\documentclass[fleqn,twoside]{article}
\usepackage{espcrc2}

\usepackage{graphicx}
\usepackage[figuresright]{rotating}

\def\dj{d\hskip-.4em\hbox{\char'26}\hskip-.1em}
\def\Dj{D\hskip-.8em\lower.4ex\hbox{\char'26}\hskip.3em}

\title{Performance of the ATLAS Precision Muon Chambers under LHC
	Operating Conditions}

\author{M.~Deile\address[address1]{ Max-Planck-Institut f\"ur Physik,
	F\"ohringer Ring 6, D-80805 M\"unchen, Germany}\thanks{Permanent address:
	CERN, CH-1211 Geneva 23, Switzerland},
	{H.~Dietl}\addressmark[address1],
	{J.~Dubbert}\address[address2]{
	Ludwig-Maximilians-Universit\"at M\"unchen, Am Coulombwall 1,
	D-85748 Garching, Germany},
	{S.~Horvat}\addressmark[address1]\thanks{Permanent address: 
	Institut Ru\dj er Bo\v skovi\' c, 10 001 Zagreb, Croatia.},
	{O.~Kortner}\addressmark[address1],
	{H.~Kroha}\addressmark[address1],
	{A.~Manz}\addressmark[address1],
	{S.~Mohrdieck}\addressmark[address1],
	{F.~Rauscher}\addressmark[address2],
	{R.~Richter}\addressmark[address1],
	{A.~Staude}\addressmark[address2]}
	       
\begin{document}


\begin{abstract}
For the muon spectrometer of the ATLAS detector at the large hadron collider
(LHC),
large drift chambers consisting of 6 to 8 layers of pressurized drift tubes
are used for precision tracking covering an active area of
5000~m$^2$ in the toroidal field of superconducting air core magnets. 
The chambers
have to provide a spatial resolution of 41 $\mu$m with Ar:CO2 (93:7) gas mixture
at an absolute pressure of
3 bar and gas gain of 2$\cdot$10$^4$. The environment in which the chambers will 
be operated is characterized by high neutron and  
$\gamma$ background with counting rates
of up to 100~s$^{-1}$cm$^{-2}$. The resolution and efficiency of a chamber from the serial
production for ATLAS has been investigated in a 100~GeV 
muon beam at photon irradiation rates as expected during LHC operation.
A silicon strip detector telescope was used as external reference in the 
beam. The spatial resolution of a chamber is degraded by 4~$\mu$m at the highest
background rate. The detection efficiency of the drift tubes is unchanged under
irradiation. A tracking efficiency of 98\% at the highest rates has been
demonstrated.
\end{abstract}

\maketitle


\section{Introduction}

The ATLAS detector is a multi-purpose detector built for the search for
the standard model Higgs boson and new physics at the large hadron collider
(LHC) at CERN. 
A striking feature of its design is the muon spectrometer, which is designed
to measure muon momenta
in the 0.4~T field of a superconduction air-core toroid magnet
with an accuracy of 3\% over a wide momentum range; at
1~TeV/c a momentum resolution of 10\% is achieved.

The muon trajectories will be measured by three
stations of monitored drift-tube (MDT) chambers. These chambers consist of two
triple- or quadruple-layers of pressurized aluminium drift tubes of 0.4~mm wall
thickness and 29.970~mm diameter which are filled with 
a gas mixtures of Ar:CO$_2$(93:7) at an absolute pressure
of 3~bar. Operated at a gas gain of 2$\cdot$10$^4$, the
tubes must provide an average spatial resolution of better than 100~$\mu$m.
In a six-layer chamber with six tubes hit, this corresponds to a spatial
resolution of the chamber of better than 100~$\mu$m/$\sqrt{6}$=41~$\mu$m.

\begin{figure*}
	\includegraphics[width=\linewidth]{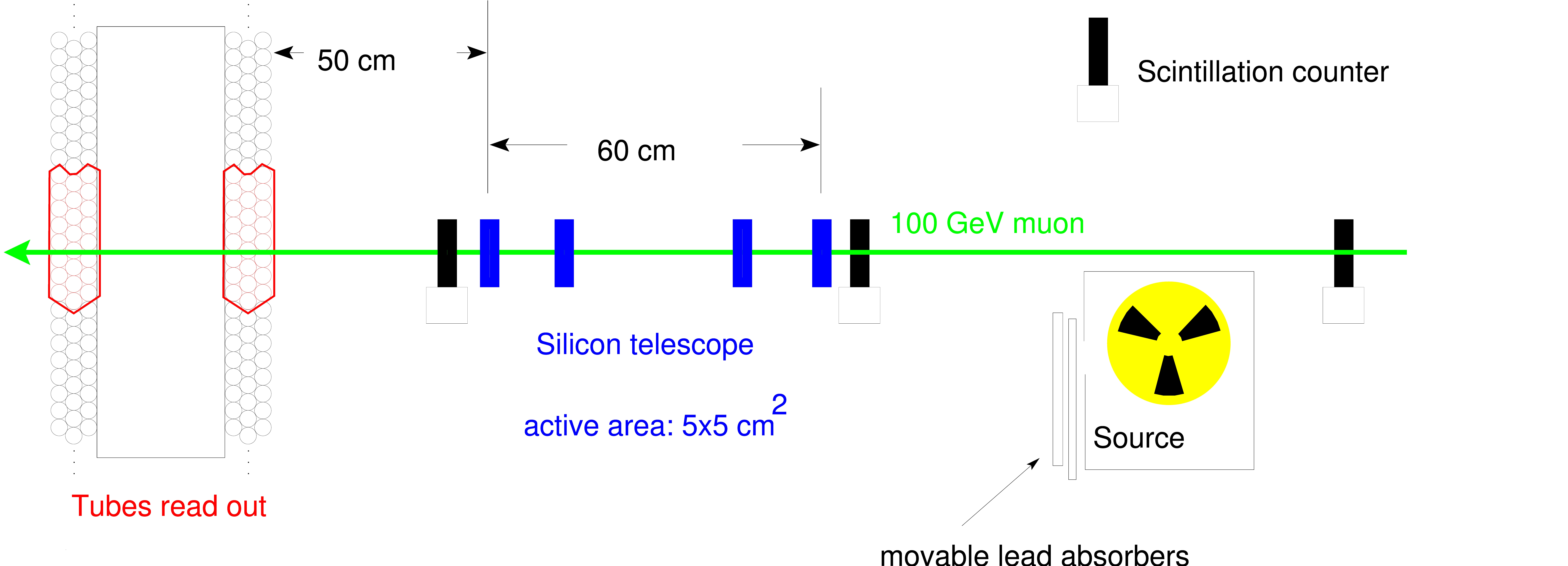}
	\caption{\label{fig1}Sketch of the experimental set-up.}
\end{figure*}

The operating conditions of the ATLAS
muon chambers at the LHC are characterized by
unprecedentedly high neutron and $\gamma$ background. The chambers will
experience background counting rates ranging from 8~s$^{-1}$cm$^{-2}$
to 100~s$^{-1}$cm$^{-2}$
\cite{TDR}. In summer 2002, one of the largest MDT
chambers constructed for the muon
spectrometer with 432 drift tubes of 3.8~m length \cite{MPI1}\cite{MPI2}
was tested in the $\gamma$ irradiation facility at CERN with a
740~GBq $^{137}$Cs source and a 100~GeV muon beam. To take into account
uncertainties in
the background estimates, the chamber was irradiated with photon
counting rates of up to 183~s$^{-1}$cm$^{-2}$. A silicon strip
tracking detector was
used as external reference 50~cm in front of the muon chamber (see
Figure~\ref{fig1}).

\section{The Spatial Resolution of the Drift Tubes}

The hits detected by the silicon detector
telescope (STEL) are used to reconstruct the
straight trajectories of the muons. The extrapolation of these trajectories into
the triple-layer of the chamber closest to the telescope determines
the impact radii $r_{STEL}$ in
the tubes traversed by the muons with an accuracy of 20~$\mu$m. The knowledge of
the impact radii $r_{STEL}$ allows a precise determination of the
space-to-drift-time relationship $r(t)$. After $r(t)$ has been determined, the
spatial resolution of a single drift tube is given by
\begin{eqnarray}
	\sigma(r_{STEL}):=\sqrt{Var\left(r(t)-r_{STEL}\right)}.
	\nonumber
\end{eqnarray}

\begin{figure*}
\begin{center}
	\includegraphics[width=\linewidth]{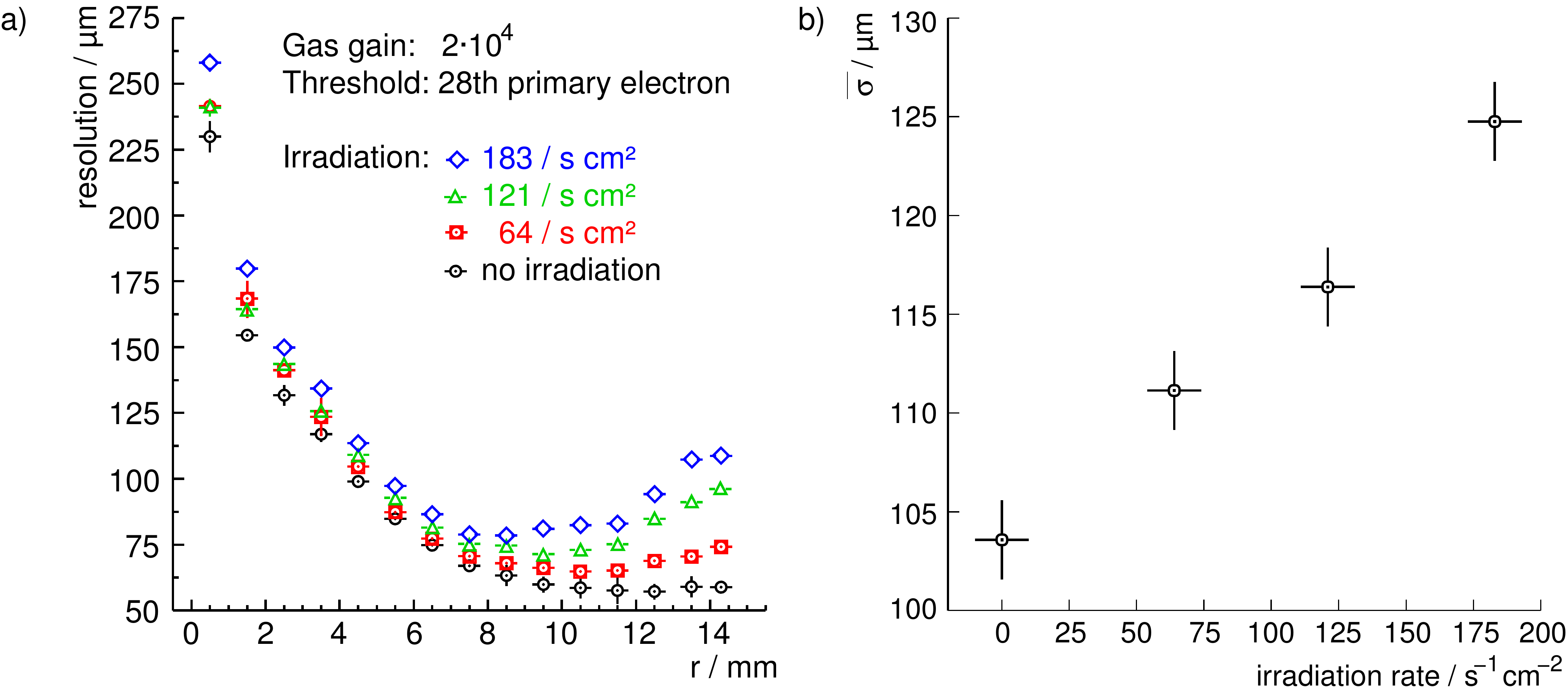}
	\caption{\label{fig2}a) Spatial resolution of a single drift tube
		as a function of the impact radius $r$ for different levels of
		irradiation. b) Average spatial resolution as a function of the irradiation level.}
	\includegraphics[width=\linewidth]{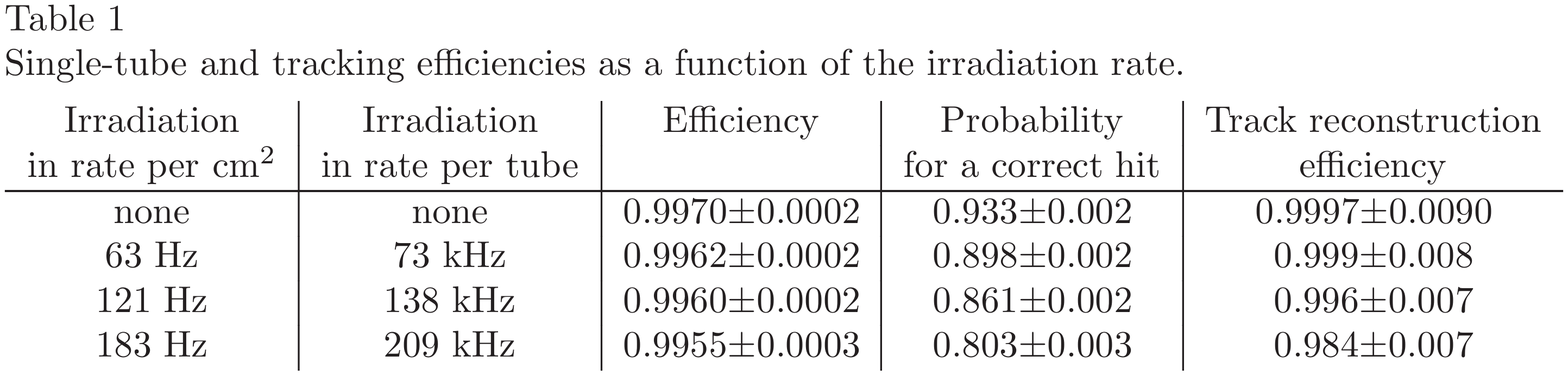}		
\end{center}
\end{figure*}

Figure~\ref{fig2}a shows the spatial resolution of a drift tube as a
function of the impact radius and the $\gamma$ irradiation rate. The spatial
resolution is degraded with increasing irradiation rate for large impact radii
because of the space charge effect.
When a photon is converted into an electron
causing a hit in a drift tube, positive ions are drifting from the avalanche
region around the wire towards the tube wall. The
space charge of the ions change the electric
field inside the tube and, hence, the drift velocity
of the ionization electrons. If the irradiation rate is low,
the muon hits in the tubes rarely overlap in time with the positive ions from
the photon conversion.
At high irradiation rates, however, the probability  increases that 
ions from background photon conversions are still present, when muons hit the
tubes. The presence of the ions causes increasing fluctuations in the
space-to-drift-time
relationship. Since for the track position measurement,
only one space-to-drift-time
relationship common to all events can be used, this leads to a decreasing
spatial resolution. Because of the longer drift time, the degradation is worse
for large impact radii. The space charge effect in drift tubes for MDT chambers
was first measured in \cite{rate_effects}. The results presented here are in
accord with the these measurements.

Figure~\ref{fig1}b) shows the average spatial resolution of the drift tubes 
defined by
\begin{eqnarray}
	\bar{\sigma} := \sqrt{\frac{1}{14.6\ \mathrm{mm}}
			\int\limits_0^{14.6\ \mathrm{mm}}\sigma^2(r)\, dr}.
	\nonumber
\end{eqnarray}
At the highest irradiation rate of 100~s$^{-1}$cm$^{-2}$ expected during
the ATLAS operation, the spatial resolution is degraded by only 10~$\mu$m
compared to the case without irradiation (see Figure~\ref{fig2}b). The average
resolution achieved with a non-irradiated chamber is slightly above
100~$\mu$m. With the test-beam data we could show
that the discriminator threshold of the read-out electronics
can be lowered from the 28th to the 25th primary ionization electron
without introducin excess noise.
This results in an improved average spatial resolution of 99~$\mu$m instead of
104~$\mu$m. A further reduction of the threshold is under study.
We also found that raising the gas gain instead is not
profitable due to the increased space-charge effect.



\section{Detection Efficiencies}

Not only the spatial resolution, but also the efficiency of a drift tube may
deteriorate under irradiation. The knowledge of the impact radii in the
drift tubes
allows for the measurement of the detection efficiency of the tubes.
The results of our measurements are summarized in Table~1.
The efficiency is always above 0.995 independently from the irradiation rate.

In the offline analysis, always the first hits after
the trigger time are selected. Therefore
the probability of chosing a wrong hit increases with the photon background 
rate. At the irradiation rate of
183~s$^{-1}$cm$^{-2}$ in the test
which corresponds to a hit rate of 209~kHz for 3.8~m tubes,
this probability is 11\%. It is less than 1 because of $\delta$~rays created in
the tube walls by the traversing muons.

Because of the redundancy of a chamber -- at least six tubes are traversed by
a muon -- a high track-reconstruction efficiency of greater than 98\% is
maintained up to the highest background rates, 
if one requires at least 3 correct hits in the muon chamber.


\section{Summary}

A large monitored drift-tube chamber was operated at LHC background conditions
and above in the gamma irradiation facility at CERN.
The measurements show that all chamber
will have a spatial resolution of better than 50~$\mu$m and that
the track reconstruction is not significantly deteriorated.


\section*{Acknowledgements}
We would like to thank S.~Zimmermann and A.~Lanza
for providing us components for the data acquisition and 
the high voltage distribution.
M. Deile thanks Brookhaven National Laboratory and the Max-Planck-Institut
f\"ur Physik in Munich for their support.


\end{document}